\documentclass[%
 reprint,
 amsmath,amssymb,
aps,
pra,
]{revtex4-1}

\usepackage{bm}
\usepackage{graphicx}

\usepackage{placeins}

\newcommand*{\la}{\langle}
\newcommand*{\ra}{\rangle}

\usepackage{dcolumn}
\newcolumntype{d}[1]{D{.}{.}{#1}} 

\begin{document}

  \title{Asymptotics of the exchange splitting energy 
for a~diatomic molecular ion~from a~volume 
integral formula of~symmetry-adapted perturbation theory}

\author{Piotr Gniewek}
\email[]{pgniewek@tiger.chem.uw.edu.pl}
\author{Bogumi\l{} Jeziorski}
 \email[]{jeziorsk@chem.uw.edu.pl}

 \affiliation{Faculty of Chemistry, University of Warsaw, 
Pasteura 1, 02-093 Warsaw, Poland}

\date{\today}

\begin{abstract}
The exchange splitting energy $J$ of the lowest \emph{gerade} and 
\emph{ungerade} states of 
the H$_2^+$ molecular ion was  calculated using a volume integral expression 
of symmetry-adapted perturbation theory and standard basis set  techniques of 
quantum chemistry. The performance of the proposed expression was compared 
to the well known surface integral formula. Both formulas involve the primitive 
function which we calculated employing either the Hirschfelder-Silbey 
perturbation theory or the conventional Rayleigh-Schr\"odinger perturbation 
theory (the polarization expansion). Our calculations show that very accurate 
values of $J$ can be obtained using the proposed volume integral formula. 
When the Hirschfelder-Silbey primitive function is used in both formulas 
the volume formula gives much more accurate results than the surface integral 
expression. We also show that using the volume integral formula 
with the primitive function approximated by 
Rayleigh-Schr\"odinger perturbation theory, one correctly obtains only the 
first four terms in the asymptotic expansion of the exchange splitting energy.  
\end{abstract}

\pacs{31.15.B-,31.10.+z}

\maketitle

\section{Introduction}

From the very advent of quantum chemistry, the exchange energy has been one 
of most prominent concepts of this scientific discipline \cite{Heitler:27}. 
It is especially important for theories of molecular binding and magnetism 
\cite{Herring:62,Herring:66}. The hydrogen molecular ion, 
H$_2^+$, is the simplest system for which exchange energy can be defined. 
In this case it is the half of the difference between the energies of the 
lowest \emph{gerade} and \emph{ungerade} states:
\begin{equation}
 J = \tfrac{1}{2} ( E_g - E_u ).
\end{equation}

Being the simplest system with a chemical bond, H$_2^+$ is a very important 
model for more complicated systems. It has served as a benchmark system for 
Symmetry Adapted Perturbation Theories (SAPT), see e.g. 
\cite{Chipman:73,Jeziorski:77}. 
It was also proposed as a model of alkali dimer cations \cite{Scott:04}.

Because the wave equation for H$_2^+$ separates in elliptic coordinates, many 
analytical results have been obtained for this system. 
Holstein and Herring \cite{Holstein:52, Herring:62} were the first 
to calculate the leading term, $-(2/e) R e^{-R}$, of the asymptotic expansion 
of $J$:
\begin{equation}\label{eq:J_invR_expansion}
 J(R) = 2 e^{-R-1} R ( j_0 + j_1\, R^{-1} + 
 j_2 \, R^{-2} + j_3 \, R^{-3} + \ldots )
\end{equation}
where $R$ is the internuclear distance. Their approach relied on calculating 
$J$ as a surface integral over the median plane $M$:
\begin{equation}\label{eq:HerringHolstein}
 J_{\textrm{surf}}[\phi] = \frac{ - \int_M \phi \mathbf{\nabla} \phi \textrm{d}
\mathbf{S} }{ \la \phi | \phi \ra  - 2 \int_{\textrm{right}} \phi^2 \textrm d V },
\end{equation}
where $\phi$ is the so-called primitive function \cite{Kutzelnigg:80}, 
which will be defined later, and ``right'' denotes the half of the whole space 
to the right of the the median plane (we use atomic units in this 
equation and throughout the paper). 
A similar calculation was included in the Landau and Lifschitz's 
textbook on quantum mechanics \cite{LandauLifschitz}. Bardsley \emph{et al.} 
\cite{Bardsley:75} used exponential parametrization of the localized function 
$\phi$ and obtained two leading terms of $J$. The third term was calculated by 
Ovchinnikov and Sukhanov \cite{Ovchinnikov:65} by means of iterative 
solution of H$_2^+$ differential equations. Komarov and Slavyanov 
\cite{Komarov:67} and Damburg and Propin \cite{Damburg:68} used 
asymptotic solutions of the ordinary differential equations for the H$_2^+$ 
wave function and 
 obtained four and nine leading terms of $J$, respectively. Brezin and 
Zinn-Justin \cite{Brezin:79} showed the connection between the leading 
term of $J$ and the large $n$ form of van der Waals $C_n$ coefficients 
of H$_2^+$. Tang \emph{et al.} \cite{Tang:91} recovered the leading term
 of $J$ by analytical summation to infinity of dominating terms of polarization
 theory. Such selective summation leads to the localized function $\phi$ 
of Herring and Holstein, as was pointed out by Scott \emph{et al.} 
\cite{Scott:91}. 

The most complete results for H$_2^+$ were obtained by \v{C}\'\i\v{z}ek 
\emph{et al.} \cite{Cizek:86}, Graffi \emph{et al.} \cite{Graffi:85}, 
and Damburg \emph{et al.} \cite{Damburg:84}, who showed that the expansion 
of energy eigenvalues of H$_2^+$ in powers of $1/R$ is Borel-summable for 
complex internuclear separations R. This Borel sum has a branch cut along the 
real $R$ axis, and taking the limit of real $R$ requires addition of explicit 
imaginary ``counter terms''. The imaginary part of the Borel sum determines 
the asymptotics of the van der Waals coefficients by a dispersion relation 
(this is a rigorous justification of Brezin and Zinn-Justin's observation 
\cite{Brezin:79}). \v{C}\'\i\v{z}ek \emph{et al.} 
\cite{Cizek:86} gave also formulas for the 
exponentially small terms, and explicit numerical values of first 52 $j_k$'s 
of the expansion (\ref{eq:J_invR_expansion}). 

Recently Burrows, Dalgarno and Cohen \cite{Burrows:10} developed 
an algebraic perturbation theory, based on asymptotic solutions of H$_2^+$ 
differential equations and comparison technique. With their method they 
obtained second, third and fourth term of (\ref{eq:J_invR_expansion}) with 
relative errors of $-$2.8\%, $-$17.8\% and 36.9\%, respectively. Nevertheless, 
no previous work has succeeded in obtaining the expansion 
(\ref{eq:J_invR_expansion}) by means of standard \emph{ab initio} approaches 
of quantum chemistry. As Whitton and Byers-Brown have pointed out 
\cite{Whitton:76}, this is partly due to the fact that 
Rayleigh-Schr\"odinger perturbation theory must be summed to infinite order
 to yield the leading term of $J$. 

The technique of Holstein and Herring was 
extended to the neutral H$_2$ molecule  in the independent 
works of Gor'kov and Pitaevskii \cite{Gorkov:64} 
and Herring and Flicker \cite{Herring:64}. Extensions to many-electron 
systems were also provided \cite{Scott:04, Bardsley:75, Smirnov:65, 
Tang:92, Jamieson:09}.

In this communication we present a  method of reproducing the asymptotic 
expansion (\ref{eq:J_invR_expansion}) using a  volume integral formula 
of symmetry-adapted  perturbation theory (SAPT). We apply our method 
to the H$_2^+$ ion to show its effectiveness for a system for which 
the exact solution is known \cite{Cizek:86, Graffi:85,Damburg:84}. 
Our procedure employs standard basis set techniques of 
electronic structure theory, therefore it generalizes straightforwardly
to many-electron systems. It is now being applied in our group to 
the H$_2$ molecule, for which the validity
of the results of Gor'kov and Pitaevskii \cite{Gorkov:64} and of  Herring 
and Flicker \cite{Herring:64}  
has been questioned \cite{Burrows:12}.

This paper is organized as follows: in Section \ref{sec:J_and_phi} we recall
the definition of the primitive function $\phi$ and derive the volume integral 
formula for the exchange energy. Section \ref{sec:phi_approximations} presents 
the approximations to $\phi$ that we use: the 
Hirschfelder-Silbey and Rayleigh-Schr\"odinger perturbation theories. 
We describe the computational aspects  of our study (the basis sets, 
extrapolation and fitting techniques) in Section 
\ref{sec:computational_aspects}. Section \ref{sec:results} describes 
the results of our investigation: the convergence with respect to the order 
of perturbation theory and with respect to the size 
of the basis set, and the accuracy of different approximations of $J$.
Our article is closed with concluding remarks in Section \ref{sec:conclusions}.

\section{\label{sec:J_and_phi}Exchange energy and the primitive function}

The derivation of the surface integral formula (\ref{eq:HerringHolstein})
 was given in Refs. \cite{Herring:62} and \cite{Bardsley:75}. Here we 
derive the volume integral formula for $J$ in terms of the primitive 
function~$\phi$. For an exhaustive analysis of the concept of primitive 
function we refer the reader to the paper by  Kutzelnigg \cite{Kutzelnigg:80}. 

The primitive function $\phi$ is defined as a  linear combination of the 
asymptotically degenerate  \emph{gerade} and \emph{ungerade} wave functions, 
$\psi_g$ and $\psi_u$,
\begin{equation}
 \phi = c_1 \psi_g + c_2 \psi_u ,
\end{equation}
which is localized on the nucleus $a$, in the sense that 
\begin{equation}\label{eq:localization_condition}
 \la \phi_0 | P_{ab} \phi \ra = o(R^{-n}),  
\end{equation}
for all $n > 0$, where $\phi_0 = 1s_a$ is the ground-state wave function of 
the hydrogen atom centered on the nucleus~$a$ and $P_{ab}$ is the operator 
of the reflection in the median plane of H$_2^+$.
Note that Kutzelnigg \cite{Kutzelnigg:80} used a more general definition and 
proposed the term {\em genuine primitive function} for the function satisfying 
the condition (\ref{eq:localization_condition}). Since we will use 
perturbation approximations to the primitive function, it is convenient 
to impose intermediate normalization:
\begin{equation}\label{eq:intermediate_normalization}
 \la \phi_0 | \phi \ra = 1 .
\end{equation}
 
Introducing the interaction energies $\mathcal E_g$ and $\mathcal E_u$,
\begin{equation}
 \mathcal E_g = E_g - E_0, \quad \quad \mathcal E_u = E_u - E_0,
\end{equation}
with $E_0 = -\frac{1}{2}$ being the ground state energy of the hydrogen atom, 
we may write the Schr\"odinger equation for the \emph{gerade} and 
\emph{ungerade} states as
\begin{equation}\label{eq:Schroedinger_g_u}\begin{aligned}
 ( H_0 - E_0 ) \psi_g &= ( \mathcal E_g - V ) \psi_g, \\ 
 ( H_0 - E_0 ) \psi_u &= ( \mathcal E_u - V ) \psi_u .
\end{aligned}\end{equation}
The unperturbed Hamiltonian $H_0$ and the interaction operator $V$ are
\begin{equation}\label{eq:H0_V_def}
 H_0 = -\frac{1}{2} \nabla^2 - \frac{1}{r_a}, 
  \quad V = -\frac{1}{r_b} + \frac{1}{R} ,
\end{equation}
where $r_a$ and $r_b$ are the distances of the electron to the nuclei $a$
 and $b$, respectively.

\begin{figure}[ht]
  \includegraphics[width=0.5\textwidth]{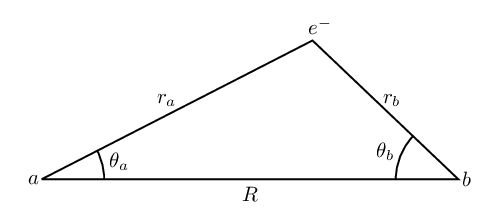}
    \caption{\label{fig:coordinates} Coordinates used in our study. 
Nuclei are denoted by $a$ and $b$.}
\end{figure}

Fig. \ref{fig:coordinates} shows coordinates that we use. The wave functions 
$\psi_g$ and $\psi_u$ are expressed through $\phi$ as
\begin{equation}\label{eq:psig_psiu_via_phi}
   c_1 \psi_g = A_g \phi, \quad \quad   c_2 \psi_u = A_u \phi ,
\end{equation}
where $A_g$ and $A_u$ are the symmetry projectors  defined as
\begin{equation}
 A_g = \tfrac{1}{2} (1 + P_{ab}), \quad A_u = \tfrac{1}{2} (1 - P_{ab}).
\end{equation}
After inserting formulas (\ref{eq:psig_psiu_via_phi}) into 
(\ref{eq:Schroedinger_g_u}) and taking inner products 
with $\phi_0$, one obtains 
\begin{equation}\begin{aligned}
 \mathcal E_g \la \phi_0 | ( 1 + P_{ab} ) \phi \ra 
    &= \la \phi_0 | V ( 1 + P_{ab} ) \phi \ra , \\
 \mathcal E_u \la \phi_0 | ( 1 - P_{ab} ) \phi \ra 
    &= \la \phi_0 | V ( 1 - P_{ab} ) \phi \ra . 
\end{aligned}\end{equation}
The solution for $J = \frac{1}{2} ( \mathcal E_g - \mathcal E_u ) 
= \frac{1}{2} ( E_g - E_u )$ is 
\begin{equation}\label{eq:volume_integral_J}
 J_{\textrm{SAPT}}[\phi] = \frac{ \la \phi_0 | V P_{ab} \phi \ra 
- \la \phi_0 | V \phi \ra \la \phi_0 | P_{ab} \phi \ra }{1 
- \la \phi_0 | P_{ab} \phi \ra^2 } .
\end{equation}
Note that this expression for $J$ contains only volume integrals and does not 
involve cancellation of long range terms --- both parts of the numerator decay 
exponentially, in accordance with (\ref{eq:localization_condition})---so that 
this expression can be used for very large $R$ 
without loss of significant figures. 
When the primitive function $\phi$ is expanded in powers of $V$, Eq. 
(\ref{eq:volume_integral_J})  generates the expansions  
of the exchange energy appearing in the symmetry-adapted perturbation theory  
\cite{Jeziorski:94,Szalewicz:05,Moszynski:07}. 
 We will refer to Eq. (\ref{eq:volume_integral_J}) as the \emph{volume integral 
formula}  or the SAPT formula for the exchange energy.

\section{\label{sec:phi_approximations}Approximations to the primitive function}

In principle $\phi$ could be obtained as a combination of variationally 
calculated $\psi_g$ and $\psi_u$ in appropriate dimer bases. This would however 
quickly lead to a loss of accuracy for large $R$. We therefore decided to test 
other approximations to $\phi$, that can be obtained directly, without the 
knowledge of $\psi_g$ and $\psi_u$.


The Hirschfelder-Silbey perturbation theory (HS) \cite{Hirschfelder:66} 
is constructed to provide a perturbation expansion  of 
the primitive function in orders of the perturbation $V$
\begin{equation} \label{HS_exp}
\phi = \phi_{\textrm{HS}}^{(0)} +\phi_{\textrm{HS}}^{(1)} 
+\phi_{\textrm{HS}}^{(2)} + \cdots .
\end{equation}
It converges to the results of  variational calculation with
the same basis set, provided that this basis set 
is invariant under symmetry operations \cite{Chalasinski:77}. 
The equations for the consecutive 
corrections $\phi_{\textrm{HS}}^{(n)}$  to the HS wave function are 
\cite{Chalasinski:77}:
\begin{equation}\label{eq:fHS-classic}\begin{aligned}
 \phi_{\textrm{HS}}^{(n)} = &- R_0 V \phi_{\textrm{HS}}^{(n-1)} 
+ \sum_{k=1}^n E_g^{(k)} R_0 A_g \phi_{\textrm{HS}}^{(n-k)} \\
           &+ \sum_{k=1}^n E_u^{(k)} R_0 A_u \phi_{\textrm{HS}}^{(n-k)} ,
\end{aligned}\end{equation}
where the energy corrections $E_g^{(n)}$ and $E_u^{(n)}$ are given by
\begin{equation}\begin{aligned}
 E_\nu^{(n)} = & \la \phi_0 | A_\nu \phi_0 \ra^{-1} \bigg ( 
\la \phi_0 | V A_\nu \phi_{\textrm{HS}}^{(n-1)} \ra  \\
 &- \sum_{k=1}^{n-1} E_\nu^{(k)} \la \phi_0 | A_\nu \phi_{\textrm{HS}}^{(n-k)} 
\ra \bigg ), \quad \nu = g, u.
\end{aligned}\end{equation}
The zeroth-order wave function and energy are those of the unperturbed hydrogen
 atom,
 $ \phi_{\textrm{HS}}^{(0)} = \phi_0 \equiv 1s_a$,  $E_g^{(0)} = E_u^{(0)}
 = E_0 \equiv -\tfrac{1}{2}$.
The resolvent $R_0$ is defined by 
\begin{equation}
 R_0 = (H_0 - E_0 + P_0)^{-1} (1 - P_0),
\end{equation}
where $P_0 = | \phi_0 \ra \la \phi_0 |$ is the operator projecting on the 
unperturbed wave function. 


The standard Rayleigh-Schr\"odinger perturbation theory 
applied to molecular interactions with $H_0$ and $V$ defined as in 
Eq. (\ref{eq:H0_V_def}) 
is known as 
the polarization expansion or polarization approximation 
\cite{Hirschfelder:67}. It gives in finite order a good 
approximation to the primitive function. Strictly speaking the polarization
 approximation gives an asymptotic representation of the primitive function 
in the following sense \cite{Jeziorski:82}:
\begin{equation}
 \phi = \sum_{k = 0}^n \phi_{\textrm{RS}}^{(k)} 
+ \mathcal O( R^{ -\kappa (n+1)}) ,
\end{equation}
with $\kappa = 2$ when at least one of interacting subsystems has a net charge, 
and $\kappa = 3$ otherwise. The wave function corrections in this theory, 
$\phi_{\textrm{RS}}^{(n)}$, are defined recursively by
\begin{equation}\label{eq:phi_pol_def}
 \phi_{\textrm{RS}}^{(n)} = - R_0 V \phi_{\textrm{RS}}^{(n-1)} 
+ \sum_{k=1}^n E_{\textrm{RS}}^{(k)} R_0 \phi_{\textrm{RS}}^{(n-k)} ,
\end{equation}
and the energy corrections $E_{\textrm{RS}}^{(n)}$ are calculated as 
$E_{\textrm{RS}}^{(n)}~=~\la \phi_0 | V \phi_{\textrm{RS}}^{(n-1)} \ra$.
 The unperturbed wave function, unperturbed energy and reduced resolvent are 
the same as in the HS theory.

\section{\label{sec:computational_aspects}Computational aspects}

Basis set used by us consists of functions 
\begin{equation}\label{eq:basis_set} 
 \chi_c^{N,M}  = \mathcal{C}_{N,M} e^{-r_c} L_{N}^{2M+2}(2 r_c) 
r_c^M P_M( \cos \theta_c) , 
\end{equation}
where $c = a,b$ and $L_N^{2M+2}(x)$ and $P_M(x)$ are the generalized Laguerre 
and Legendre polynomials, respectively, 
defined as in e.g. Ref.~\cite{Abramowitz:65}. 
The normalization constant of the basis function $\chi_c^{N,M}$ 
is denoted by $\mathcal{C}_{N,M}$.   The angles 
$\theta_a$ and $\theta_b$ are the interior ones of the triangle given 
by $r_a$, $r_b$, $R$ (see Fig. \ref{fig:coordinates}), so that
 $\theta_b = P_{ab} \theta_a$ and $ \chi_b^{N,M} = P_{ab} \chi_a^{N,M}$. This
 convention for $\theta_b$ was used by Bardsley \emph{et al}. 
in Ref. \cite{Bardsley:75}. 
Two center integrals generated when using this 
  basis  set are easily  calculated using the conventional elliptic coordinates 
$\xi = (r_a+r_b)/R$ 
and $\eta = (r_a-r_b)/R$. The unperturbed wave function is explicitly included 
in the basis, $\phi_0 = \chi_a^{0,0}$. 

Basis functions centered on the same nucleus are 
orthonormal, whereas overlap integrals of functions centered on different 
atoms decay exponentially,
\begin{equation}
 \la \chi_a^{N_1,M_1} | \chi_b^{N_2,M_2} \ra \sim e^{-R} .
\end{equation}
This reduces linear dependencies in the basis set at large $R$, 
allowing for accurate calculations of the asymptotic constants $j_k$. 
The values of $N$ and $M$ are the 
same for basis functions centered on nucleus $a$ and $b$, therefore basis 
(\ref{eq:basis_set}) is invariant under the action 
of $P_{ab}$, and converged 
HS theory gives results exact in this basis 
\cite{Chalasinski:77}. 
We introduce a hierarchy of basis sets through the parameter $\Omega$ 
constraining $N$ and $M$:
\begin{equation}\label{eq:def_Omega}
 N+M \leq \Omega .
\end{equation}
This hierarchy is useful for making extrapolations to the complete basis set 
limit. $N$ and $M$ are treated symmetrically in 
Eq. (\ref{eq:def_Omega}) in order to maintain consistency with the 
multipole expansion of the wave function, and to provide
the best convergence at large $R$.

The basis set (\ref{eq:basis_set}) is appropriate for large internuclear 
distances $R$ but is inadequate for small ones because of strong linear 
dependencies appearing at larger values of  $\Omega$.   We decided that 
the smallest internuclear distance used in the fitting of the asymptotic 
constants $j_k$ is $R = 60$. For this distance the octupole precision 
(exact to 64 significant decimal digits) was required  to perform accurate 
calculations for $\Omega=25$  (702 basis functions). 

Chipman and Hirschfelder used basis similar to (\ref{eq:basis_set}), but with 
monomials in $r_a$ and $r_b$ instead of Laguerre polynomials, when they applied 
different perturbation theories to H$_2^+$ \cite{Chipman:73}. 
The basis (\ref{eq:basis_set}) restricted to functions centered on the nucleus 
$a$ was used by Coulson \cite{Coulson:41} and by Morgan and Simon 
\cite{Morgan:80} in their calculations of van der Waals 
coefficients of H$_2^+$.

The regularity of the $\Omega$-dependence of the computed values of $J$ permits 
an efficient application of extrapolation technique to accelerate basis set 
convergence. We used Levin's $u$-transformation  of the form \cite{Levin:72}:
\begin{equation}
 U_n = \frac{ \sum_{i=0}^n (-1)^i \binom{ n }{ i } (i+1)^{n-2}
  Z_i  A_i^{-1} }{ \sum_{i=0}^n (-1)^i \binom{ n }{ i } (i+1)^{n-2}
  A_i^{-1} } ,
\end{equation}
where $U_n$ is the resulting accelerated sequence, and 
$Z_i = A_0 + A_1 + \ldots + A_i$ is the partial sum to be accelerated. 
The Levin $u$-transformation is considered to be the best general purpose 
convergence accelerating sequence transformation \cite{Press:07}. 
For an efficient and numerically stable algorithm and general 
discussion of this and similar transformations see Ref.~\cite{Weniger:89}.

In   case of basis extrapolation there are many possible choices of $A_i$ 
and $Z_i$.  After extensive analysis of the performance of different choices 
we  decided to report   results obtained with 
the 6-term Levin $u$-transformation 
applied to the 6 best basis sets. With  this choice we have 
 $Z_n$=$J$($\Omega$=$n$ $+$20),   $A_0$= $J$($\Omega$= 20), 
and $A_n$= $J$($\Omega$=$n$ $+$20) $-$ $J$($\Omega$=$n$ $+$19)  for $n$ $>$0.
 
We used the least squares method to extract the asymptotic constants $j_k$ 
from the calculated values of $J(R)$. In order to increase 
the numerical stability of our analysis, we scaled  the values of  $J(R) $ 
 multiplying them by $e^{R+1}/(2R)$ prior 
to the fitting procedure. The fitting functions were then   polynomials 
in $R^{-1}$, in accordance with Eq. (\ref{eq:J_invR_expansion}):
\begin{equation}
 f(R) = \sum_{i=0}^L \frac{ \tilde{j_i} }{ R^i } .
\end{equation}
It is important to choose the appropriate degree of the fitting polynomial $L$. 
A fit with too small $L$ would fail to extract all the available information 
from the  calculated values while a  too large $L$ would lead to inaccurate 
results. 

In our calculations we used a grid of 46 equally spaced values of internuclear 
distance  $R$ = 60, 62, \ldots, 150 in the fitting procedure. We used an 
additional ``test set'' of 9 internuclear distances $R$ = 65, 75, \ldots 145 
to assess the quality of fits. Analysis of the errors given by fits with 
different $L$ for the 9 points in the test set allowed us to determine the 
optimal values of $L$. We found that the optimal value of $L$ is 10 when the 
volume integral formula is used.
For the surface integral expression the optimal choice of $L$ is 5.

\section{\label{sec:results}Results and discussion}

\subsection{Convergence of perturbation theory}

When the primitive function is approximated by either $\phi_{\textrm{HS}}$ 
or $\phi_{\textrm{RS}}$, the exchange energy $J$ can be expanded in powers 
of $V$,
\begin{equation}\label{eq:series_Jn}
 J = \sum_{k = 1}^\infty J_{\textrm{SAPT}}^{(k)}[\phi] ,
\end{equation}
and the corrections $J_{\textrm{SAPT}}^{(n)}[\phi]$ are given by
\begin{widetext}
\begin{equation}\label{eq:term_Jn}
 J_{\textrm{SAPT}}^{(n)} [\phi]  =  
 \la \phi_0 | V P_{ab} \phi^{(n-1)} \ra - \sum_{k=0}^{n-1}
 \la \phi_0 | V \phi^{(k)} \ra \la \phi_0 | P_{ab} \phi^{(n-k-1)} \ra 
 + \mathcal O(e^{-2R}) ,
\end{equation}
\end{widetext}
where $\phi$ stands either for $\phi_{\textrm{HS}}$ or $\phi_{\textrm{RS}}$. 
 
\begin{figure}[ht]
  \includegraphics[width=0.5\textwidth]{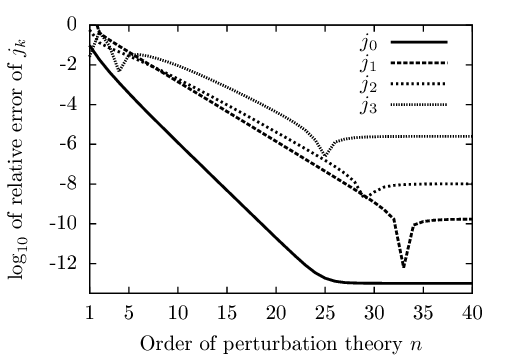}
    \caption{\label{fig:jk_vs_npt_HS} Convergence of $j_0$, $j_1$, $j_2$ 
and $j_3$ as a function of the perturbation order $n$ of the 
Hirschfelder-Silbey perturbation theory. Volume integral formula and basis 
set $\Omega=25$ were used.
Errors are calculated with respect to the exact values given in 
Ref. \cite{Cizek:86}. The errors remaining beyond
 the 30th order are due to the basis set incompleteness. }
\end{figure}

The Hirschfelder-Silbey perturbation theory is characterized by very good 
convergence  \cite{Chalasinski:77, Cwiok1994symmetry}. We observed that the 
convergence radius for the series of exchange corrections 
$J_{\textrm{SAPT}}^{(n)} [\phi_{\textrm{HS}}]$ was close to 2 and 
was almost independent on the internuclear distance $R$. 
These convergence properties result in a similarly good convergence 
of the asymptotic coefficients  $j_k$ of Eq. (\ref{eq:J_invR_expansion}) 
fitted to the results of calculations for finite $R$.
Fig.~\ref{fig:jk_vs_npt_HS} demonstrates the convergence of $j_k$'s
obtained from the HS theory.

\begin{figure}[ht]
  \includegraphics[width=0.5\textwidth]{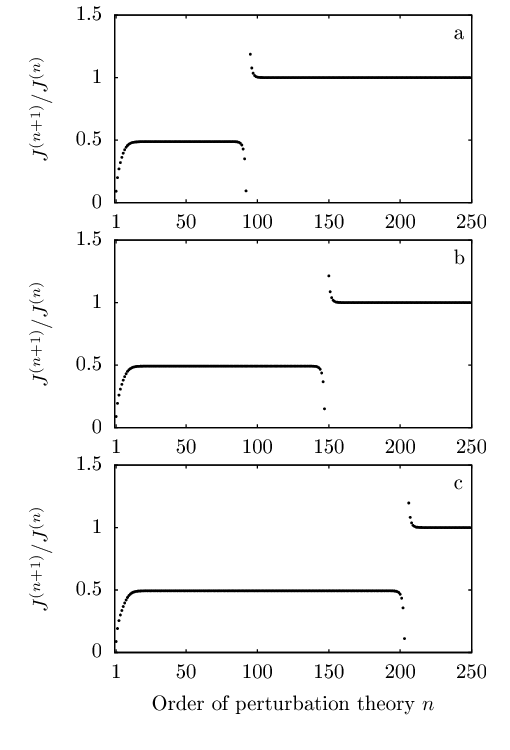}
    \caption{\label{fig:RS_ratios} Fractions of subsequent corrections to 
the exchange energy $J_{\textrm{SAPT}}^{(n)}[\phi_{\textrm{RS}}]$ for three 
different internuclear distances: a) $R$ = 40, b) $R$ = 60, 
c) $R$ = 80. Basis $\Omega=25$ was used. }
\end{figure}

In comparison to the HS theory, the convergence properties of the 
Rayleigh-Schr\"odinger perturbation expansion are much more complicated 
\cite{Cwiok:92:Pol}. This is reflected in the convergence 
of the exchange energy corrections 
$J_{\textrm{SAPT}}^{(n)} [\phi_{\textrm{RS}}]$ calculated 
from Eq. (\ref{eq:term_Jn}). These corrections are identical with those 
of the Symmetrized Rayleigh-Schr\"odinger perturbation theory (SRS) 
\cite{Jeziorski:78}. 
For perturbation orders $n$ larger than 10 and smaller than some critical 
value $n_{\textrm{crit}}$, the  ratios of exchange energy corrections 
$J_{\textrm{SAPT}}^{(n+1)} [\phi_{\textrm{RS}}]
 / J_{\textrm{SAPT}}^{(n)} [\phi_{\textrm{RS}}]$ are approximately
 equal to 0.5.  For $n$ larger than $n_{\textrm{crit}}$ 
 these ratios are close to 1. 
Value of $n_{\textrm{crit}}$ increases with internuclear distance $R$. 
This is illustrated in Fig. \ref{fig:RS_ratios}.

It is clear that the convergence properties of the series of corrections 
$J_{\textrm{SAPT}}^{(n)} [\phi_{\textrm{RS}}]$ are pathological and 
it is not possible to obtain the exact limit of the series (\ref{eq:term_Jn}) 
with $\phi=\phi_{\textrm{RS}}$ through direct summation. The practical limit 
is obtained when corrections are summed up to $n_{\textrm{crit}}$. This method 
of summation gives very accurate values of $J(R)$ and the figure illustrating 
the convergence of the fitted asymptotic constants $j_k$, $k=0,1,2,3$, would 
be practically indistinguishable from Fig.  \ref{fig:jk_vs_npt_HS} 
illustrating the convergence of the HS theory. 

We calculated the convergence radius $\rho$ of the series of corrections 
$\phi_{\textrm{RS}}^{(n)}$  and found that it is always greater than 
1 but only marginally at large $R$. This convergence radius is determined 
\cite{Cwiok:92:Pol} by a pair of branch points of the two lowest 
lying eigenvalues of the non-hermitian operator
$H_0 + \zeta V$, where $\zeta$ is a complex 
scaling parameter. The radius $\rho$ can be written in the form $1+\gamma$, 
with $\gamma \sim e^{-2R}$ (for instance  $\gamma = 1.84 \cdot 10^{-47}$ 
for $R=60$ and  $\gamma = 1.90 \cdot 10^{-124}$ for $R=150$). 
The physical value of the scaling parameter, 
 $\zeta=1$, lies  therefore inside the convergence circle
of the $\phi_{\textrm{RS}}^{(n)}$ series.
Thus, the series of exchange corrections 
$J_{\textrm{RS}}^{(n)}$ must converge despite the apparent stabilization 
of the high-order terms. Since the sum of corrections  
$\phi_{\textrm{RS}}^{(n)}$ satisfies the Schr\"odinger equation 
the polarization series converges  to the exact, \emph{gerade} wave function 
of H$_2^+$ satisfying $P_{ab} \phi_{\textrm{RS}} = \phi_{\textrm{RS}}$ 
in the limit $n\rightarrow \infty$. Thus, in view of the symmetry condition 
$P_{ab} \phi_{\textrm{RS}} = \phi_{\textrm{RS}}$, the volume integral 
formula~(\ref{eq:volume_integral_J}) exhibits $0/0$ singularity at  
$n\rightarrow \infty$.

We shall now show that this singularity is removable. 
Our derivation is based on the ideas given in 
Refs. \cite{Cwiok:92:SRS}.
The limit of the series of Eq.~(\ref{eq:series_Jn}) 
with $\phi=\phi_{\textrm{RS}}$
can be obtained from the limit $\zeta \rightarrow 1$ in  
Eq.~(\ref{eq:volume_integral_J}) in which $V$ and $\phi$ are replaced by
$\zeta V$ and $\phi_{\textrm{RS}}(\zeta)$, respectively,
\begin{widetext}
\begin{equation}\label{eq:limit_zeta_1}
 J_{\textrm{SAPT}}[\phi_{\textrm{RS}}] = \lim_{\zeta \rightarrow 1} 
 \frac{ 
 \la \phi_0 | \zeta V P_{ab} \phi_{\textrm{RS}}(\zeta) \ra
 \la \phi_0 |  \phi_{\textrm{RS}}(\zeta) \ra -
 \la \phi_0 | \zeta V \phi_{\textrm{RS}}(\zeta) \ra
 \la \phi_0 | P_{ab} \phi_{\textrm{RS}}(\zeta) \ra 
 }{
 \la \phi_0 | \phi_{\textrm{RS}}(\zeta) \ra^2 -
 \la \phi_0 | P_{ab} \phi_{\textrm{RS}}(\zeta) \ra^2 
 } .
\end{equation}
\end{widetext}
Note that we use here a slight modification of the 
volume integral formula (\ref{eq:volume_integral_J}) which 
is independent of the normalization of $\phi$.
The limit in Eq.~(\ref{eq:limit_zeta_1}) can be 
obtained with the use of the l'Hospital rule.
The derivative of the numerator $\mathcal{N}$
of the left hand side of Eq.~(\ref{eq:limit_zeta_1}) 
is
\begin{equation}\label{eq:Jzeta_derivative_num}\begin{aligned}
 \frac{ d \mathcal{N} }{ d \zeta } \bigg |_{\zeta=1} 
 &=
 \la \phi_0 | V \psi_g \ra \la \phi_0 | (1-P_{ab}) \psi^{(1)} \ra + \\
 &- \la \phi_0 | V (1-P_{ab}) \psi^{(1)} \ra \la \phi_0 | \psi_g \ra ,
\end{aligned}\end{equation}
where $\psi^{(1)}$ is the derivative of $\phi_{\textrm{RS}}(\zeta)$ 
with respect to $\zeta$ at $\zeta=1$,
\begin{equation}\label{eq:phiRS_zeta_derivative}
 \psi^{(1)} =
 \frac{ d \phi_{\textrm{RS}} }{ d \zeta } \bigg |_{\zeta=1}
 = - \sum_{s \neq g} \frac{ \la \psi_s | V \psi_g \ra }{ E_s - E_g } \psi_s ,
\end{equation}
where the summation involves all excited states $s$ of $H$ 
(the energy and wavefunction of an excited state $s$ are denoted by 
$E_s$ and $\psi_s$, respectively).
Eq.~(\ref{eq:Jzeta_derivative_num}) can be rearranged to yield
\begin{equation}\begin{aligned}
 \frac{ d \mathcal{N} }{ d \zeta } \bigg |_{\zeta=1}  & =
 \la \phi_0 | \psi_g \ra \la \phi_0 | (P_{ab} -1 ) (H - E_g) \psi^{(1)} \ra = \\
 & = \la \phi_0 | \psi_g \ra \la \phi_0 | (1-P_{ab}) V \psi_g \ra .
\end{aligned}\end{equation}
The $\zeta$ derivative of the denominator $\mathcal D$ of Eq.~(\ref{eq:limit_zeta_1})
reads
\begin{equation}
  \frac{ d \mathcal{D} }{ d \zeta } \bigg |_{\zeta=1}  =
 2 \la \phi_0 | \psi_g \ra \la \phi_0 | (1 - P_{ab}) \psi^{(1)} \ra .
\end{equation}
The contribution of $\psi_u$ dominates in Eq.~(\ref{eq:phiRS_zeta_derivative}),
therefore
\begin{equation}
  \frac{ d \mathcal{D} }{ d \zeta } \bigg |_{\zeta=1}  =
 \frac{ \la \phi_0 | \psi_g \ra }{ J }  \big [
 2 \la \phi_0 | \psi_u \ra  \la \psi_u | V \psi_g \ra 
 + \mathcal O(e^{-R})
 \big ] .
\end{equation}
Consequently the limit of Eq.~(\ref{eq:limit_zeta_1}) is 
\begin{equation}
 J_{\textrm{SAPT}}[\phi_{\textrm{RS}}] = J \frac{
 \la \phi_0 | (1 - P_{ab} ) V \psi_g \ra
 }{
 2 \la \phi_0 | \psi_u \ra \la \psi_u | V \psi_g \ra 
 }
 + \mathcal O(e^{-2R}) .
\end{equation} 
Expressing $\psi_g$ and $\psi_u$ via the primitive function
$\phi$ we find
\begin{equation}\label{eq:Jsapt_phiRS_c}\begin{aligned}
 &J_{\textrm{SAPT}} [\phi_{\textrm{RS}}] =  \\
 &\quad=J \la \phi | \phi \ra 
 \frac{ 
 \la (1-P_{ab}) \phi_0 | V (1+P_{ab}) \phi \ra 
 }{
 \la (1-P_{ab}) \phi | V (1+P_{ab}) \phi \ra 
 } + \mathcal O(J^2) .
\end{aligned}\end{equation}
The Symmetrized Rayleigh-Schr\"odinger perturbation theory
is therefore convergent, albeit it gives a limit different from the 
true value of the exchange energy. This limit is nevertheless
asymptotically exact: when the primitive function $\phi$ in
Eq.~(\ref{eq:Jsapt_phiRS_c}) is approximated using the 
multipole expansion \cite{Jeziorski:82}, one obtains the following
expression:
\begin{equation}
 \frac{ J_{\textrm{SAPT}} [\phi_{\textrm{RS}}] }{J}  =  
1 + \frac{w_4}{R^4} + \frac{w_5}{R^5} 
 + \frac{w_6}{R^6} + \frac{w_7}{R^7} + 
 \mathcal O(R^{-8}) ,
\end{equation}
with $w_4=w_5=67/8$, $w_6=173/4$, $w_7=14657/32$.
The numerical results given in Sec. \ref{sec:volume} 
confirm this asymptotic behavior of the SRS exchange energy.

When a very similar reasoning is applied to the removable singularity of 
$J_{\textrm{surf}}[\phi_{\textrm{RS}}]$ one obtains
\begin{equation}
 \frac{ J_{\textrm{surf}}[\phi_{\textrm{RS}}] }{J} 
 = 1 + \mathcal O (e^{-R}) ,
\end{equation}
in agreement with the conclusions of Ref. \cite{Scott:93}.

 \subsection{Basis set convergence\label{subsec:basis_convergence}}

\begin{figure}[ht]
  \includegraphics[width=0.5\textwidth]{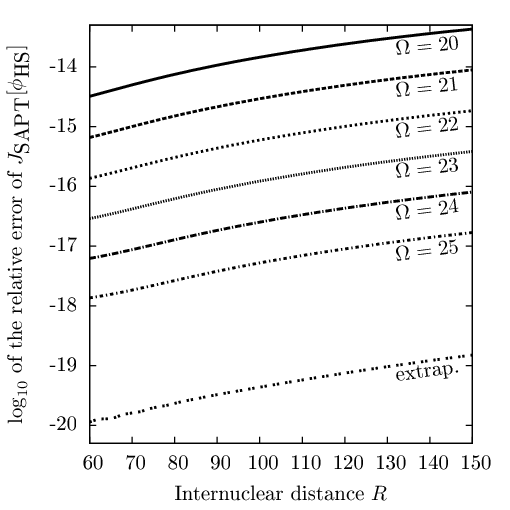}
    \caption{\label{fig:basis_convergence} Decimal logarithms of 
errors of $J_{\textrm{SAPT}}^{(n)} [\phi_{\textrm{HS}}]$ and values 
extrapolated from 6 best bases: $\Omega$=20,\ldots,25. Errors are calculated 
with respect to exact results 
of \v{C}\'\i\v{z}ek \emph{et. al.} \cite{Cizek:86}.}
\end{figure}

We found that the convergence of results with respect to $\Omega$ is 
very regular for a wide range of internuclear distances. 
When $\Omega$ is increased by 3, the relative errors of exchange energy 
(compared to exact results of \v{C}\'\i\v{z}ek \emph{et. al} \cite{Cizek:86}) 
decrease by two orders of magnitude. This behavior is shown in Fig. 
\ref{fig:basis_convergence}.

Fig. \ref{fig:basis_convergence}  also shows   that this good convergence can 
be further  improved by the  application of the Levin $u$-transformation. 
When this transformation is applied to the results calculated in basis sets 
$\Omega$=17,18,\ldots,22, one obtains exchange energy with the accuracy 
almost identical to that of $\Omega$=25. Therefore it can be 
estimated that values computed as the transformation of six energies 
$J$($\Omega$ $=$20),\ldots,$J$($\Omega$ $=$25) are of accuracy similar to that possible 
to calculate with basis $\Omega$ $=$ 28 (bases $\Omega$ $=$ 25 and $\Omega$ $=$ 28
contain 702 and 870 basis functions, respectively). 

\phantom{\ref{tab:J_HS_RS_surf}}

\subsection{Comparison of numerical results obtained from the volume 
and surface integral formulas\label{sec:volume}}

\begin{figure}[ht]
  \includegraphics[width=0.5\textwidth]{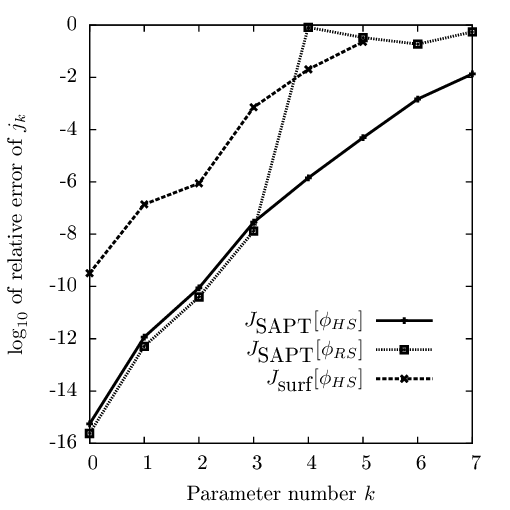}
    \caption{\label{fig:Jsapt} Decimal logarithms of errors 
of $j_k$, $k = 0, \ldots 7$ obtained with the primitive functions  
$\phi_{\textrm{HS}}$ and  $\phi_{\textrm{RS}}$. Results extrapolated from 
basis sets $\Omega$=20,21,\ldots,25.}
\end{figure}

In Fig. \ref{fig:Jsapt} we show errors of the asymptotic expansion parameters 
$j_k$ for the two considered 
approximations to the primitive function $\phi$. 
These parameters are also given in Table \ref{tab:J_HS_RS_surf}.

\begin{table*}[t]
\caption{\label{tab:J_HS_RS_surf} Values of $j_k$ obtained from different
approximations and the exact values calculated by \v{C}\'\i\v{z}ek \emph{et al.} 
\cite{Cizek:86}. The degrees of the fitting polynomials were 8 for 
$J_{\textrm{SAPT}}$ and 4 for $J_{\textrm{surf}}$. The values extrapolated from 
the best six basis sets ($\Omega$=20,\ldots,25) were used in the fits. }
\begin{ruledtabular}
\begin{tabular}{ c d{3.12} d{3.24} d{3.24} d{3.16} }
 &  \multicolumn{1}{c}{ $J_{\textrm{exact}}$ }
 &  \multicolumn{1}{c}{ $J_{\textrm{SAPT}}[\phi_{\textrm{HS}}]$ } 
 &  \multicolumn{1}{c}{ $J_{\textrm{SAPT}}[\phi_{\textrm{RS}}]$ } 
 &  \multicolumn{1}{c}{ $J_{\textrm{surf}}[\phi_{\textrm{HS}}]$ }  \\
\hline 
$j_0$ & 
  -1 & 
  -0.999\ 999\ 999\ 999\ 999\ 45 & 
  -0.999\ 999\ 999\ 999\ 999\ 76 & 
  -0.999\ 999\ 999\ 68   \\ 
 $j_1$ & 
  -0.5 & 
  -0.500\ 000\ 000\ 000\ 58 & 
  -0.500\ 000\ 000\ 000\ 26 & 
  -0.500\ 000\ 069   \\ 
 $j_2$ & 
   3.125 & 
   3.125\ 000\ 000\ 28 & 
   3.125\ 000\ 000\ 12 & 
   3.124\ 997\ 2  \\ 
 $j_3$ & 
   2.729\ 166\ 67 & 
   2.729\ 166\ 59 & 
   2.729\ 166\ 63 & 
   2.731\ 1  \\ 
 $j_4$ & 
  10.216\ 146  & 
  10.216\ 160 & 
   1.8 & 
  10.01  \\ 
 $j_5$ & 
  37.864\ 3  & 
  37.862\ 5 & 
  25. & 
  47.  \\ 
 $j_6$ & 
  113.26 & 
  113.43 & 
   92. & 
  \multicolumn{1}{l}{ \phantom{aaa} --} \\
 $j_7$ & 
  789.2  & 
  778.5 & 
  353.  &  
 \multicolumn{1}{l}{ \phantom{aaa} --} \\
\end{tabular}
\end{ruledtabular}
\end{table*}

 It can be seen that the volume integral formula with $\phi_{\textrm{HS}}$
 is able to reproduce all  $j_k$ of eq. (\ref{eq:J_invR_expansion}), provided 
sufficient basis set and numerical precision are used. It is also clear from 
  Fig. \ref{fig:Jsapt} that the   SRS theory is able to reproduce only four 
leading coefficients of the asymptotic expansion (\ref{eq:J_invR_expansion}).

We found by least square fitting that the relative error of 
the SRS exchange energy with respect to the HS one is well 
represented by a polynomial in $1/R$:
\begin{equation}
 \frac{ J_{\textrm{SAPT}}[\phi_{\textrm{RS}}] }{ 
J_{\textrm{SAPT}}[\phi_{\textrm{HS}}] } = 1 + \frac{\tilde w_4}{R^4} 
+ \frac{\tilde w_5}{R^5} + \frac{\tilde w_6}{R^6} + \frac{\tilde w_7}{R^7} + \ldots
\end{equation}
with $\tilde w_4$ = 8.375 000 000 000(3), $\tilde w_5$ = 8.375 000 000(3), 
$\tilde w_6$ = 43.250 000(1), $\tilde w_7$ = 458.031 3(3) (the numbers in parentheses give 
the uncertainties of the respective last reported digits). The values obtained by fitting are
in perfect agreement with the ones calculated from Eq.~(\ref{eq:Jsapt_phiRS_c}).

Somewhat surprisingly we found that $\phi_{\textrm{HS}}$ and 
$\phi_{\textrm{RS}}$   give practically identical results
 (up to more than 20 digits)   when used in the surface integral formula 
(\ref{eq:HerringHolstein}). For instance when $R = 100$ and $\Omega = 25$ 
we obtained  $  J_{\textrm{surf}}[\phi_{\textrm{HS}}] 
=J_{\textrm{surf}}[\phi_{\textrm{RS}}] =
-2.749 \ 901 \ 239 \ 50  \  \cdot 10^{-42}$  
 while the exact value is  
$J_{\textrm{exact}}= 
-2.749 \ 901 \ 239 \ 63  \ \cdot 10^{-42}$.  
 The errors of these  approximate values are however much larger  than 
those obtained with the volume integral and the HS primitive function. 
We found that  the relative errors of   
$J_{\textrm{SAPT}}[\phi_{\textrm{HS}}]$,  defined as 
$\Delta J = |(J-J_{\textrm{exact}})/J_{\textrm{exact}}|$,  
are of the order of 10$^{-17}$ while the 
relative errors of $  J_{\textrm{surf}}[\phi_{\textrm{HS}}] 
=J_{\textrm{surf}}[\phi_{\textrm{RS}}]$ range from 10$^{-13}$ to 10$^{-9}$
 for $R$ between 60 and 150  (the values of $J_{\rm exact}$ are calculated
 from the exact asymptotic constants  \cite{Cizek:86}).  
The errors of  $J_{\textrm{SAPT}}[\phi_{\textrm{RS}}]$ are of the order
 of 10$^{-6}$--10$^{-8}$ in this range of distances.  These increased errors 
(compared to those of  $J_{\textrm{SAPT}}[\phi_{\textrm{HS}}]$) are not related
to a remaining basis set incompleteness but are  
caused by the incorrect values of the higher $j_k$ coefficients predicted by
 $J_{\textrm{SAPT}}[\phi_{\textrm{RS}}]$.

The higher errors resulting from using the surface integral formula 
can be understood when the quality of the wave function 
is considered. Accuracy of $J_{\textrm{surf}}[\phi]$ 
depends mainly on the accuracy of the wave function $\phi$ 
in the vicinity of the median plane $M$. We can inspect the quality 
of any approximate wave function $\psi$ by analyzing the local 
energy associated with this wave function
\begin{equation}
 E_{\textrm{loc}}(\bm r) = \frac{H \psi (\bm r)}{ \psi(\bm r)} .
\end{equation}
The local energy was used in a similar context by Bartlett who applied it 
to assess 
the quality of his numerical approximation to the wave function of
the helium atom in Ref. \cite{Bartlett:55}. 


\begin{figure}[h]
  \includegraphics[width=0.5\textwidth]{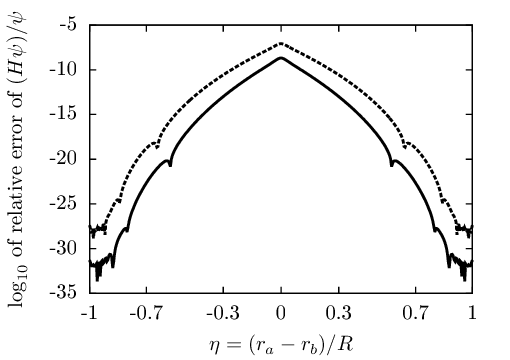}
    \caption{\label{fig:loc_en_bond} Errors of the local energy 
$(H \psi) / \psi$ for $\psi = (1+P_{ab}) \phi_{\textrm{HS}}$ calculated 
for the line joining the nuclei ($\xi = 1$, $\eta = -1,\ldots,1$). 
Internuclear distance $R$ = 100, basis sets $\Omega=20$ (dashed line) 
and $\Omega=25$ (solid line).}
\end{figure}

Fig. \ref{fig:loc_en_bond}  visualizes the errors of $E_{\textrm{loc}}$ 
for the line joining the nuclei. The reference energy was calculated as 
the sum of $E_g^{(n)} (\textrm{HS})$ up to $n=150$. This graph presents 
$E_{\textrm{loc}}$ for the symmetrized $\phi_{\textrm{HS}}$ function, 
$\psi =(1+P_{ab}) \phi_{\textrm{HS}}$, obtained
with  the $\Omega=25$ basis set. The graph for $\phi_{\textrm{RS}}$ 
is not given, as it would be indistinguishable from the one for 
$\phi_{\textrm{HS}}$ with this scale. It can be seen  that the the primitive 
functions investigated in our work give a very good description of the part 
of configuration space close to the nuclei, but have much larger errors 
a the median plane, i.e. in the region relevant for the accuracy 
of the surface integral formula. 

\section{\label{sec:conclusions}Conclusions}

The SAPT volume integral formula and the surface integral formula give
very accurate values of the exchange splitting energy  
when the primitive function is approximated either by
the Hirschfelder-Silbey or the Rayleigh-Schr\"odinger perturbation theories.
However, the volume integral expression exhibits much superior basis set 
convergence than the surface integral one. This is due to the fact that 
the accuracy of the latter 
depends strongly on the quality of the 
wave function (and thus the basis set) at the median plane $M$, i.e. far from 
the nuclei. The very good (and very similar)
basis set  convergence of  $J_{\textrm{SAPT}}[\phi_{\textrm{HS}}]$ and 
$J_{\textrm{SAPT}}[\phi_{\textrm{RS}}]$ 
is  further improved by extrapolation techniques such as the 
Levin $u$-transformation. 
We also found that the values the asymptotic constants obtained from
$J_{\textrm{surf}}[\phi_{\textrm{HS}}]$ and
$J_{\textrm{surf}}[\phi_{\textrm{RS}}]$ are almost identical.

We find it remarkable that the perturbation expansion of 
$J_{\textrm{SAPT}}[\phi_{\textrm{RS}}]$, which is 
equivalent to the Symmetrized Rayleigh-Schr\"odinger perturbation theory (SRS),
does converge but gives accurately only the first 
four terms of the asymptotic 
expansion of Eq. (\ref{eq:J_invR_expansion}). 
The unphysical values of further terms are  
due to the removable 0/0 singularity 
of $J_{\textrm{SAPT}}[\phi_{\textrm{RS}}]$ 
in the limit $n \rightarrow \infty$.

It should be pointed out that application of the proposed 
volume integral formula 
involves standard basis set and integral evaluation techniques of 
quantum chemistry and therefore  this expression can
be more easily employed in case of many-electron systems, both ionic 
and neutral, than the surface integral one.

\begin{acknowledgments}
The authors acknowledge discussions with 
Prof.~R.~Moszy\'nski and Mr.~M.~Lesiuk and thank
Dr.~T.~Korona for reading and commenting on
the manuscript. 
This work was supported by the Polish Ministry of Science, 
grant no.~NN204~182840.
\end{acknowledgments}

\bibliography{ref}

\begin{thebibliography}{46}%
\makeatletter
\providecommand \@ifxundefined [1]{%
 \@ifx{#1\undefined}
}%
\providecommand \@ifnum [1]{%
 \ifnum #1\expandafter \@firstoftwo
 \else \expandafter \@secondoftwo
 \fi
}%
\providecommand \@ifx [1]{%
 \ifx #1\expandafter \@firstoftwo
 \else \expandafter \@secondoftwo
 \fi
}%
\providecommand \natexlab [1]{#1}%
\providecommand \enquote  [1]{``#1''}%
\providecommand \bibnamefont  [1]{#1}%
\providecommand \bibfnamefont [1]{#1}%
\providecommand \citenamefont [1]{#1}%
\providecommand \href@noop [0]{\@secondoftwo}%
\providecommand \href [0]{\begingroup \@sanitize@url \@href}%
\providecommand \@href[1]{\@@startlink{#1}\@@href}%
\providecommand \@@href[1]{\endgroup#1\@@endlink}%
\providecommand \@sanitize@url [0]{\catcode `\\12\catcode `\$12\catcode
  `\&12\catcode `\#12\catcode `\^12\catcode `\_12\catcode `\%12\relax}%
\providecommand \@@startlink[1]{}%
\providecommand \@@endlink[0]{}%
\providecommand \url  [0]{\begingroup\@sanitize@url \@url }%
\providecommand \@url [1]{\endgroup\@href {#1}{\urlprefix }}%
\providecommand \urlprefix  [0]{URL }%
\providecommand \Eprint [0]{\href }%
\providecommand \doibase [0]{http://dx.doi.org/}%
\providecommand \selectlanguage [0]{\@gobble}%
\providecommand \bibinfo  [0]{\@secondoftwo}%
\providecommand \bibfield  [0]{\@secondoftwo}%
\providecommand \translation [1]{[#1]}%
\providecommand \BibitemOpen [0]{}%
\providecommand \bibitemStop [0]{}%
\providecommand \bibitemNoStop [0]{.\EOS\space}%
\providecommand \EOS [0]{\spacefactor3000\relax}%
\providecommand \BibitemShut  [1]{\csname bibitem#1\endcsname}%
\let\auto@bib@innerbib\@empty
\bibitem [{\citenamefont {Heitler}\ and\ \citenamefont
  {London}(1927)}]{Heitler:27}%
  \BibitemOpen
  \bibfield  {author} {\bibinfo {author} {\bibfnamefont {W.}~\bibnamefont
  {Heitler}}\ and\ \bibinfo {author} {\bibfnamefont {F.}~\bibnamefont
  {London}},\ }\href {\doibase 10.1007/BF01397394} {\bibfield  {journal}
  {\bibinfo  {journal} {Z. Phys.}\ }\textbf {\bibinfo {volume} {44}},\ \bibinfo
  {pages} {455} (\bibinfo {year} {1927})}\BibitemShut {NoStop}%
\bibitem [{\citenamefont {Herring}(1962)}]{Herring:62}%
  \BibitemOpen
  \bibfield  {author} {\bibinfo {author} {\bibfnamefont {C.}~\bibnamefont
  {Herring}},\ }\href {\doibase 10.1103/RevModPhys.34.631} {\bibfield
  {journal} {\bibinfo  {journal} {Rev. Mod. Phys.}\ }\textbf {\bibinfo {volume}
  {34}},\ \bibinfo {pages} {631} (\bibinfo {year} {1962})}\BibitemShut
  {NoStop}%
\bibitem [{\citenamefont {Herring}(1966)}]{Herring:66}%
  \BibitemOpen
  \bibfield  {author} {\bibinfo {author} {\bibfnamefont {C.}~\bibnamefont
  {Herring}},\ }in\ \href@noop {} {\emph {\bibinfo {booktitle} {{Magnetism: A
  Treatise on Modern Theory and Materials}}}},\ Vol.~\bibinfo {volume} {IV},\
  \bibinfo {editor} {edited by\ \bibinfo {editor} {\bibfnamefont
  {G.}~\bibnamefont {Rado}}\ and\ \bibinfo {editor} {\bibfnamefont
  {H.}~\bibnamefont {Suhl}}}\ (\bibinfo  {publisher} {Academic Press},\
  \bibinfo {address} {New York},\ \bibinfo {year} {1966})\BibitemShut {NoStop}%
\bibitem [{\citenamefont {Chipman}\ and\ \citenamefont
  {Hirschfelder}(1973)}]{Chipman:73}%
  \BibitemOpen
  \bibfield  {author} {\bibinfo {author} {\bibfnamefont {D.~M.}\ \bibnamefont
  {Chipman}}\ and\ \bibinfo {author} {\bibfnamefont {J.~O.}\ \bibnamefont
  {Hirschfelder}},\ }\href {\doibase 10.1063/1.1680416} {\bibfield  {journal}
  {\bibinfo  {journal} {J. Chem. Phys.}\ }\textbf {\bibinfo {volume} {59}},\
  \bibinfo {pages} {2838} (\bibinfo {year} {1973})}\BibitemShut {NoStop}%
\bibitem [{\citenamefont {Jeziorski}\ and\ \citenamefont
  {Ko\l{}os}(1977)}]{Jeziorski:77}%
  \BibitemOpen
  \bibfield  {author} {\bibinfo {author} {\bibfnamefont {B.}~\bibnamefont
  {Jeziorski}}\ and\ \bibinfo {author} {\bibfnamefont {W.}~\bibnamefont
  {Ko\l{}os}},\ }\href@noop {} {\bibfield  {journal} {\bibinfo  {journal} {Int.
  J. Quant. Chem.}\ }\textbf {\bibinfo {volume} {12}},\ \bibinfo {pages} {91}
  (\bibinfo {year} {1977})}\BibitemShut {NoStop}%
\bibitem [{\citenamefont {{T. C. Scott, M. Aubert-Fr\'econ, G. Hadinger, D.
  Andrae, J. Grotendorst, and J. D. Morgan III}}(2004)}]{Scott:04}%
  \BibitemOpen
  \bibfield  {author} {\bibinfo {author} {\bibnamefont {{T. C. Scott, M.
  Aubert-Fr\'econ, G. Hadinger, D. Andrae, J. Grotendorst, and J. D. Morgan
  III}}},\ }\href {\doibase 10.1088/0953-4075/37/22/005} {\bibfield  {journal}
  {\bibinfo  {journal} {J. Phys. B}\ }\textbf {\bibinfo {volume} {37}},\
  \bibinfo {pages} {4451} (\bibinfo {year} {2004})}\BibitemShut {NoStop}%
\bibitem [{\citenamefont {Holstein}(1952)}]{Holstein:52}%
  \BibitemOpen
  \bibfield  {author} {\bibinfo {author} {\bibfnamefont {T.}~\bibnamefont
  {Holstein}},\ }\href {\doibase 10.1021/j150499a004} {\bibfield  {journal}
  {\bibinfo  {journal} {J. Phys. Chem.}\ }\textbf {\bibinfo {volume} {56}},\
  \bibinfo {pages} {832} (\bibinfo {year} {1952})}\BibitemShut {NoStop}%
\bibitem [{\citenamefont {Kutzelnigg}(1980)}]{Kutzelnigg:80}%
  \BibitemOpen
  \bibfield  {author} {\bibinfo {author} {\bibfnamefont {W.}~\bibnamefont
  {Kutzelnigg}},\ }\href {\doibase 10.1063/1.439880} {\bibfield  {journal}
  {\bibinfo  {journal} {J. Chem. Phys.}\ }\textbf {\bibinfo {volume} {73}},\
  \bibinfo {pages} {343} (\bibinfo {year} {1980})}\BibitemShut {NoStop}%
\bibitem [{\citenamefont {Landau}\ and\ \citenamefont
  {Lifshitz}(1965)}]{LandauLifschitz}%
  \BibitemOpen
  \bibfield  {author} {\bibinfo {author} {\bibfnamefont {L.}~\bibnamefont
  {Landau}}\ and\ \bibinfo {author} {\bibfnamefont {E.~M.}\ \bibnamefont
  {Lifshitz}},\ }\href@noop {} {\emph {\bibinfo {title} {Quantum Mechanics:
  Non-Relativistic Theory}}}\ (\bibinfo  {publisher} {Pergamon Press},\
  \bibinfo {address} {Oxford},\ \bibinfo {year} {1965})\ p.\ \bibinfo {pages}
  {291}\BibitemShut {NoStop}%
\bibitem [{\citenamefont {Bardsley}\ \emph {et~al.}(1975)\citenamefont
  {Bardsley}, \citenamefont {Holstein}, \citenamefont {Junker},\ and\
  \citenamefont {{Swati Sinha}}}]{Bardsley:75}%
  \BibitemOpen
  \bibfield  {author} {\bibinfo {author} {\bibfnamefont {J.~N.}\ \bibnamefont
  {Bardsley}}, \bibinfo {author} {\bibfnamefont {T.}~\bibnamefont {Holstein}},
  \bibinfo {author} {\bibfnamefont {B.~R.}\ \bibnamefont {Junker}}, \ and\
  \bibinfo {author} {\bibnamefont {{Swati Sinha}}},\ }\href {\doibase
  10.1103/PhysRevA.11.1911} {\bibfield  {journal} {\bibinfo  {journal} {Phys.
  Rev. A}\ }\textbf {\bibinfo {volume} {11}},\ \bibinfo {pages} {1911}
  (\bibinfo {year} {1975})}\BibitemShut {NoStop}%
\bibitem [{\citenamefont {Ovchinnikov}\ and\ \citenamefont
  {Sukhanov}(1965)}]{Ovchinnikov:65}%
  \BibitemOpen
  \bibfield  {author} {\bibinfo {author} {\bibfnamefont {A.~A.}\ \bibnamefont
  {Ovchinnikov}}\ and\ \bibinfo {author} {\bibfnamefont {A.~D.}\ \bibnamefont
  {Sukhanov}},\ }\href@noop {} {\bibfield  {journal} {\bibinfo  {journal} {Sov.
  Phys. Dokl.}\ }\textbf {\bibinfo {volume} {9}},\ \bibinfo {pages} {685}
  (\bibinfo {year} {1965})}\BibitemShut {NoStop}%
\bibitem [{\citenamefont {{I. V. Komarov and S. Yu.
  Slavyanov}}(1967)}]{Komarov:67}%
  \BibitemOpen
  \bibfield  {author} {\bibinfo {author} {\bibnamefont {{I. V. Komarov and S.
  Yu. Slavyanov}}},\ }\href@noop {} {\bibfield  {journal} {\bibinfo  {journal}
  {Sov. Phys. {JETP}}\ }\textbf {\bibinfo {volume} {25}},\ \bibinfo {pages}
  {910} (\bibinfo {year} {1967})}\BibitemShut {NoStop}%
\bibitem [{\citenamefont {{R. J. Damburg and R. Kh.
  Propin}}(1968)}]{Damburg:68}%
  \BibitemOpen
  \bibfield  {author} {\bibinfo {author} {\bibnamefont {{R. J. Damburg and R.
  Kh. Propin}}},\ }\href {\doibase 10.1088/0022-3700/1/4/319} {\bibfield
  {journal} {\bibinfo  {journal} {J. Phys. B}\ }\textbf {\bibinfo {volume}
  {1}},\ \bibinfo {pages} {681} (\bibinfo {year} {1968})}\BibitemShut {NoStop}%
\bibitem [{\citenamefont {{E. Brezin and J. Zinn-Justin}}(1979)}]{Brezin:79}%
  \BibitemOpen
  \bibfield  {author} {\bibinfo {author} {\bibnamefont {{E. Brezin and J.
  Zinn-Justin}}},\ }\href {\doibase 10.1051/jphyslet:019790040019051100}
  {\bibfield  {journal} {\bibinfo  {journal} {J. Physique Lett.}\ }\textbf
  {\bibinfo {volume} {40}},\ \bibinfo {pages} {511} (\bibinfo {year}
  {1979})}\BibitemShut {NoStop}%
\bibitem [{\citenamefont {Tang}\ \emph {et~al.}(1991)\citenamefont {Tang},
  \citenamefont {Toennies},\ and\ \citenamefont {Yiu}}]{Tang:91}%
  \BibitemOpen
  \bibfield  {author} {\bibinfo {author} {\bibfnamefont {K.~T.}\ \bibnamefont
  {Tang}}, \bibinfo {author} {\bibfnamefont {J.~P.}\ \bibnamefont {Toennies}},
  \ and\ \bibinfo {author} {\bibfnamefont {C.~L.}\ \bibnamefont {Yiu}},\ }\href
  {\doibase 10.1063/1.460211} {\bibfield  {journal} {\bibinfo  {journal} {J.
  Chem. Phys.}\ }\textbf {\bibinfo {volume} {94}},\ \bibinfo {pages} {7266}
  (\bibinfo {year} {1991})}\BibitemShut {NoStop}%
\bibitem [{\citenamefont {{T. C. Scott, A. Dalgarno, and J. D. Morgan
  III}}(1991)}]{Scott:91}%
  \BibitemOpen
  \bibfield  {author} {\bibinfo {author} {\bibnamefont {{T. C. Scott, A.
  Dalgarno, and J. D. Morgan III}}},\ }\href {\doibase
  10.1103/PhysRevLett.67.1419} {\bibfield  {journal} {\bibinfo  {journal}
  {Phys. Rev. Lett.}\ }\textbf {\bibinfo {volume} {67}},\ \bibinfo {pages}
  {1419} (\bibinfo {year} {1991})}\BibitemShut {NoStop}%
\bibitem [{\citenamefont {{J. \v{C}\'\i\v{z}ek, R. J. Damburg, S. Graffi, V.
  Grecchi, E. M. Harrell II, J. G. Harris, S. Nakai, J. Paldus, R. Kh. Propin,
  and H. J. Silverstone}}(1986)}]{Cizek:86}%
  \BibitemOpen
  \bibfield  {author} {\bibinfo {author} {\bibnamefont {{J. \v{C}\'\i\v{z}ek,
  R. J. Damburg, S. Graffi, V. Grecchi, E. M. Harrell II, J. G. Harris, S.
  Nakai, J. Paldus, R. Kh. Propin, and H. J. Silverstone}}},\ }\href {\doibase
  10.1103/PhysRevA.33.12} {\bibfield  {journal} {\bibinfo  {journal} {Phys.
  Rev. A}\ }\textbf {\bibinfo {volume} {33}},\ \bibinfo {pages} {12} (\bibinfo
  {year} {1986})}\BibitemShut {NoStop}%
\bibitem [{\citenamefont {{S. Graffi, V. Grecchi, E. M. Harrell II, and H. J.
  Silverstone}}(1985)}]{Graffi:85}%
  \BibitemOpen
  \bibfield  {author} {\bibinfo {author} {\bibnamefont {{S. Graffi, V. Grecchi,
  E. M. Harrell II, and H. J. Silverstone}}},\ }\href {\doibase
  10.1016/0003-4916(85)90305-7} {\bibfield  {journal} {\bibinfo  {journal}
  {Ann. Phys. (N.Y.)}\ }\textbf {\bibinfo {volume} {165}},\ \bibinfo {pages}
  {441} (\bibinfo {year} {1985})}\BibitemShut {NoStop}%
\bibitem [{\citenamefont {{R. J. Damburg, R. Kh. Propin, S. Graffi, V. Grecchi,
  E. M. Harrell II, J. \v{C}\'\i\v{z}ek, J. Paldus, and H. J.
  Silverstone}}(1984)}]{Damburg:84}%
  \BibitemOpen
  \bibfield  {author} {\bibinfo {author} {\bibnamefont {{R. J. Damburg, R. Kh.
  Propin, S. Graffi, V. Grecchi, E. M. Harrell II, J. \v{C}\'\i\v{z}ek, J.
  Paldus, and H. J. Silverstone}}},\ }\href {\doibase
  10.1103/PhysRevLett.52.1112} {\bibfield  {journal} {\bibinfo  {journal}
  {Phys. Rev. Lett.}\ }\textbf {\bibinfo {volume} {52}},\ \bibinfo {pages}
  {1112} (\bibinfo {year} {1984})}\BibitemShut {NoStop}%
\bibitem [{\citenamefont {Burrows}\ \emph {et~al.}(2010)\citenamefont
  {Burrows}, \citenamefont {Dalgarno},\ and\ \citenamefont
  {Cohen}}]{Burrows:10}%
  \BibitemOpen
  \bibfield  {author} {\bibinfo {author} {\bibfnamefont {B.~L.}\ \bibnamefont
  {Burrows}}, \bibinfo {author} {\bibfnamefont {A.}~\bibnamefont {Dalgarno}}, \
  and\ \bibinfo {author} {\bibfnamefont {M.}~\bibnamefont {Cohen}},\ }\href
  {\doibase 10.1103/PhysRevA.81.042508} {\bibfield  {journal} {\bibinfo
  {journal} {Phys. Rev. A}\ }\textbf {\bibinfo {volume} {81}},\ \bibinfo
  {pages} {042508} (\bibinfo {year} {2010})}\BibitemShut {NoStop}%
\bibitem [{\citenamefont {Whitton}\ and\ \citenamefont
  {Byers-Brown}(1976)}]{Whitton:76}%
  \BibitemOpen
  \bibfield  {author} {\bibinfo {author} {\bibfnamefont {W.~N.}\ \bibnamefont
  {Whitton}}\ and\ \bibinfo {author} {\bibfnamefont {W.}~\bibnamefont
  {Byers-Brown}},\ }\href {\doibase 10.1002/qua.560100107} {\bibfield
  {journal} {\bibinfo  {journal} {Int. J. Quant. Chem.}\ }\textbf {\bibinfo
  {volume} {10}},\ \bibinfo {pages} {71} (\bibinfo {year} {1976})}\BibitemShut
  {NoStop}%
\bibitem [{\citenamefont {Gor'kov}\ and\ \citenamefont
  {Pitaevskii}(1964)}]{Gorkov:64}%
  \BibitemOpen
  \bibfield  {author} {\bibinfo {author} {\bibfnamefont {L.~P.}\ \bibnamefont
  {Gor'kov}}\ and\ \bibinfo {author} {\bibfnamefont {L.~P.}\ \bibnamefont
  {Pitaevskii}},\ }\href@noop {} {\bibfield  {journal} {\bibinfo  {journal}
  {Sov. Phys. Dokl.}\ }\textbf {\bibinfo {volume} {8}},\ \bibinfo {pages} {788}
  (\bibinfo {year} {1964})}\BibitemShut {NoStop}%
\bibitem [{\citenamefont {Herring}\ and\ \citenamefont
  {Flicker}(1964)}]{Herring:64}%
  \BibitemOpen
  \bibfield  {author} {\bibinfo {author} {\bibfnamefont {C.}~\bibnamefont
  {Herring}}\ and\ \bibinfo {author} {\bibfnamefont {M.}~\bibnamefont
  {Flicker}},\ }\href {\doibase 10.1103/PhysRev.134.A362} {\bibfield  {journal}
  {\bibinfo  {journal} {Phys. Rev.}\ }\textbf {\bibinfo {volume} {134}},\
  \bibinfo {pages} {A362} (\bibinfo {year} {1964})}\BibitemShut {NoStop}%
\bibitem [{\citenamefont {Smirnov}\ and\ \citenamefont
  {Chibisov}(1965)}]{Smirnov:65}%
  \BibitemOpen
  \bibfield  {author} {\bibinfo {author} {\bibfnamefont {B.~M.}\ \bibnamefont
  {Smirnov}}\ and\ \bibinfo {author} {\bibfnamefont {M.~I.}\ \bibnamefont
  {Chibisov}},\ }\href@noop {} {\bibfield  {journal} {\bibinfo  {journal} {Sov.
  Phys. JETP}\ }\textbf {\bibinfo {volume} {21}},\ \bibinfo {pages} {624}
  (\bibinfo {year} {1965})}\BibitemShut {NoStop}%
\bibitem [{\citenamefont {Tang}\ \emph {et~al.}(1992)\citenamefont {Tang},
  \citenamefont {Toennies}, \citenamefont {Wanschura},\ and\ \citenamefont
  {Yiu}}]{Tang:92}%
  \BibitemOpen
  \bibfield  {author} {\bibinfo {author} {\bibfnamefont {K.~T.}\ \bibnamefont
  {Tang}}, \bibinfo {author} {\bibfnamefont {J.~P.}\ \bibnamefont {Toennies}},
  \bibinfo {author} {\bibfnamefont {M.}~\bibnamefont {Wanschura}}, \ and\
  \bibinfo {author} {\bibfnamefont {C.~L.}\ \bibnamefont {Yiu}},\ }\href
  {\doibase 10.1103/PhysRevA.46.3746} {\bibfield  {journal} {\bibinfo
  {journal} {Phys. Rev. A}\ }\textbf {\bibinfo {volume} {46}},\ \bibinfo
  {pages} {3746} (\bibinfo {year} {1992})}\BibitemShut {NoStop}%
\bibitem [{\citenamefont {Jamieson}\ \emph {et~al.}(2009)\citenamefont
  {Jamieson}, \citenamefont {Dalgarno}, \citenamefont {Aymar},\ and\
  \citenamefont {Tharamel}}]{Jamieson:09}%
  \BibitemOpen
  \bibfield  {author} {\bibinfo {author} {\bibfnamefont {M.~J.}\ \bibnamefont
  {Jamieson}}, \bibinfo {author} {\bibfnamefont {A.}~\bibnamefont {Dalgarno}},
  \bibinfo {author} {\bibfnamefont {M.}~\bibnamefont {Aymar}}, \ and\ \bibinfo
  {author} {\bibfnamefont {J.}~\bibnamefont {Tharamel}},\ }\href {\doibase
  10.1088/0953-4075/42/9/095203} {\bibfield  {journal} {\bibinfo  {journal} {J.
  Phys. B}\ }\textbf {\bibinfo {volume} {42}},\ \bibinfo {pages} {095203}
  (\bibinfo {year} {2009})}\BibitemShut {NoStop}%
\bibitem [{\citenamefont {Burrows}\ \emph {et~al.}(2012)\citenamefont
  {Burrows}, \citenamefont {Dalgarno},\ and\ \citenamefont
  {Cohen}}]{Burrows:12}%
  \BibitemOpen
  \bibfield  {author} {\bibinfo {author} {\bibfnamefont {B.~L.}\ \bibnamefont
  {Burrows}}, \bibinfo {author} {\bibfnamefont {A.}~\bibnamefont {Dalgarno}}, \
  and\ \bibinfo {author} {\bibfnamefont {M.}~\bibnamefont {Cohen}},\ }\href
  {\doibase 10.1103/PhysRevA.86.052525} {\bibfield  {journal} {\bibinfo
  {journal} {Phys. Rev. A}\ }\textbf {\bibinfo {volume} {86}},\ \bibinfo
  {pages} {052525} (\bibinfo {year} {2012})}\BibitemShut {NoStop}%
\bibitem [{\citenamefont {Jeziorski}\ \emph {et~al.}(1994)\citenamefont
  {Jeziorski}, \citenamefont {Moszy\'nski},\ and\ \citenamefont
  {Szalewicz}}]{Jeziorski:94}%
  \BibitemOpen
  \bibfield  {author} {\bibinfo {author} {\bibfnamefont {B.}~\bibnamefont
  {Jeziorski}}, \bibinfo {author} {\bibfnamefont {R.}~\bibnamefont
  {Moszy\'nski}}, \ and\ \bibinfo {author} {\bibfnamefont {K.}~\bibnamefont
  {Szalewicz}},\ }\href {\doibase 10.1021/cr00031a008} {\bibfield  {journal}
  {\bibinfo  {journal} {Chem. Rev.}\ }\textbf {\bibinfo {volume} {94}},\
  \bibinfo {pages} {1887} (\bibinfo {year} {1994})}\BibitemShut {NoStop}%
\bibitem [{\citenamefont {Szalewicz}\ \emph {et~al.}(2005)\citenamefont
  {Szalewicz}, \citenamefont {Patkowski},\ and\ \citenamefont
  {Jeziorski}}]{Szalewicz:05}%
  \BibitemOpen
  \bibfield  {author} {\bibinfo {author} {\bibfnamefont {K.}~\bibnamefont
  {Szalewicz}}, \bibinfo {author} {\bibfnamefont {K.}~\bibnamefont
  {Patkowski}}, \ and\ \bibinfo {author} {\bibfnamefont {B.}~\bibnamefont
  {Jeziorski}},\ }in\ \href@noop {} {\emph {\bibinfo {booktitle}
  {Intermolecular Forces and Clusters (Structure and Bonding, volume 116)}}},\
  \bibinfo {editor} {edited by\ \bibinfo {editor} {\bibfnamefont {D.~J.}\
  \bibnamefont {Wales}}}\ (\bibinfo  {publisher} {Springer-Verlag},\ \bibinfo
  {address} {Heidelberg},\ \bibinfo {year} {2005})\ pp.\ \bibinfo {pages}
  {43--117}\BibitemShut {NoStop}%
\bibitem [{\citenamefont {{R. Moszy\'nski}}(2007)}]{Moszynski:07}%
  \BibitemOpen
  \bibfield  {author} {\bibinfo {author} {\bibnamefont {{R. Moszy\'nski}}},\
  }in\ \href@noop {} {\emph {\bibinfo {booktitle} {{Challenges and Advances in
  Computational Chemistry and Physics}}}},\ Vol.~\bibinfo {volume} {4},\
  \bibinfo {editor} {edited by\ \bibinfo {editor} {\bibfnamefont {A.~W.}\
  \bibnamefont {Sokalski}}}\ (\bibinfo  {publisher} {Springer},\ \bibinfo
  {address} {Dordrecht},\ \bibinfo {year} {2007})\ pp.\ \bibinfo {pages}
  {1--152}\BibitemShut {NoStop}%
\bibitem [{\citenamefont {Hirschfelder}\ and\ \citenamefont
  {Silbey}(1966)}]{Hirschfelder:66}%
  \BibitemOpen
  \bibfield  {author} {\bibinfo {author} {\bibfnamefont {J.~O.}\ \bibnamefont
  {Hirschfelder}}\ and\ \bibinfo {author} {\bibfnamefont {R.}~\bibnamefont
  {Silbey}},\ }\href {\doibase 10.1063/1.1727907} {\bibfield  {journal}
  {\bibinfo  {journal} {J. Chem. Phys.}\ }\textbf {\bibinfo {volume} {45}},\
  \bibinfo {pages} {2188} (\bibinfo {year} {1966})}\BibitemShut {NoStop}%
\bibitem [{\citenamefont {Cha\l{}asi\'nski}\ \emph {et~al.}(1977)\citenamefont
  {Cha\l{}asi\'nski}, \citenamefont {Jeziorski},\ and\ \citenamefont
  {Szalewicz}}]{Chalasinski:77}%
  \BibitemOpen
  \bibfield  {author} {\bibinfo {author} {\bibfnamefont {G.}~\bibnamefont
  {Cha\l{}asi\'nski}}, \bibinfo {author} {\bibfnamefont {B.}~\bibnamefont
  {Jeziorski}}, \ and\ \bibinfo {author} {\bibfnamefont {K.}~\bibnamefont
  {Szalewicz}},\ }\href {\doibase 10.1002/qua.560110205} {\bibfield  {journal}
  {\bibinfo  {journal} {Int. J. Quant. Chem.}\ }\textbf {\bibinfo {volume}
  {11}},\ \bibinfo {pages} {247} (\bibinfo {year} {1977})}\BibitemShut
  {NoStop}%
\bibitem [{\citenamefont {Hirschfelder}(1967)}]{Hirschfelder:67}%
  \BibitemOpen
  \bibfield  {author} {\bibinfo {author} {\bibfnamefont {J.~O.}\ \bibnamefont
  {Hirschfelder}},\ }\href {\doibase 10.1016/0009-2614(67)80007-1} {\bibfield
  {journal} {\bibinfo  {journal} {Chem. Phys. Lett.}\ }\textbf {\bibinfo
  {volume} {1}},\ \bibinfo {pages} {325 } (\bibinfo {year} {1967})}\BibitemShut
  {NoStop}%
\bibitem [{\citenamefont {Jeziorski}\ and\ \citenamefont
  {Ko\l{}os}(1982)}]{Jeziorski:82}%
  \BibitemOpen
  \bibfield  {author} {\bibinfo {author} {\bibfnamefont {B.}~\bibnamefont
  {Jeziorski}}\ and\ \bibinfo {author} {\bibfnamefont {W.}~\bibnamefont
  {Ko\l{}os}},\ }in\ \href@noop {} {\emph {\bibinfo {booktitle} {Molecular
  interactions}}},\ Vol.~\bibinfo {volume} {3},\ \bibinfo {editor} {edited by\
  \bibinfo {editor} {\bibfnamefont {H.}~\bibnamefont {Ratajczak}}\ and\
  \bibinfo {editor} {\bibfnamefont {W.~J.}\ \bibnamefont {Orville-Thomas}}}\
  (\bibinfo  {publisher} {Wiley},\ \bibinfo {address} {New York},\ \bibinfo
  {year} {1982})\ pp.\ \bibinfo {pages} {1--46}\BibitemShut {NoStop}%
\bibitem [{\citenamefont {Abramowitz}\ and\ \citenamefont
  {Stegun}(1965)}]{Abramowitz:65}%
  \BibitemOpen
  \bibfield  {author} {\bibinfo {author} {\bibfnamefont {M.}~\bibnamefont
  {Abramowitz}}\ and\ \bibinfo {author} {\bibfnamefont {I.~A.}\ \bibnamefont
  {Stegun}},\ }\href@noop {} {\emph {\bibinfo {title} {Handbook of mathematical
  function: with formulas, graphs and mathematical tables}}}\ (\bibinfo
  {publisher} {Dover Publications},\ \bibinfo {address} {New York},\ \bibinfo
  {year} {1965})\BibitemShut {NoStop}%
\bibitem [{\citenamefont {Coulson}(1941)}]{Coulson:41}%
  \BibitemOpen
  \bibfield  {author} {\bibinfo {author} {\bibfnamefont {C.~A.}\ \bibnamefont
  {Coulson}},\ }\href {\doibase 10.1017/S0080454100006038} {\bibfield
  {journal} {\bibinfo  {journal} {Proc. R. Soc. Edin. A-MA}\ }\textbf {\bibinfo
  {volume} {61}},\ \bibinfo {pages} {20} (\bibinfo {year} {1941})}\BibitemShut
  {NoStop}%
\bibitem [{\citenamefont {{J. D. Morgan III and B. Simon}}(1980)}]{Morgan:80}%
  \BibitemOpen
  \bibfield  {author} {\bibinfo {author} {\bibnamefont {{J. D. Morgan III and
  B. Simon}}},\ }\href {\doibase 10.1002/qua.560170609} {\bibfield  {journal}
  {\bibinfo  {journal} {Int. J. Quant. Chem.}\ }\textbf {\bibinfo {volume}
  {17}},\ \bibinfo {pages} {1143} (\bibinfo {year} {1980})}\BibitemShut
  {NoStop}%
\bibitem [{\citenamefont {Levin}(1972)}]{Levin:72}%
  \BibitemOpen
  \bibfield  {author} {\bibinfo {author} {\bibfnamefont {D.}~\bibnamefont
  {Levin}},\ }\href {\doibase 10.1080/00207167308803075} {\bibfield  {journal}
  {\bibinfo  {journal} {Int. J. Comput. Math.}\ }\textbf {\bibinfo {volume}
  {3}},\ \bibinfo {pages} {371} (\bibinfo {year} {1972})}\BibitemShut {NoStop}%
\bibitem [{\citenamefont {Press}\ \emph {et~al.}(2007)\citenamefont {Press},
  \citenamefont {Teukolsky}, \citenamefont {Vetterling},\ and\ \citenamefont
  {Flannery}}]{Press:07}%
  \BibitemOpen
  \bibfield  {author} {\bibinfo {author} {\bibfnamefont {W.~H.}\ \bibnamefont
  {Press}}, \bibinfo {author} {\bibfnamefont {S.~A.}\ \bibnamefont
  {Teukolsky}}, \bibinfo {author} {\bibfnamefont {W.~T.}\ \bibnamefont
  {Vetterling}}, \ and\ \bibinfo {author} {\bibfnamefont {B.~P.}\ \bibnamefont
  {Flannery}},\ }\href@noop {} {\emph {\bibinfo {title} {Numerical Recipes 3rd
  Edition: The Art of Scientific Computing}}}\ (\bibinfo  {publisher}
  {Cambridge University Press},\ \bibinfo {address} {New York},\ \bibinfo
  {year} {2007})\BibitemShut {NoStop}%
\bibitem [{\citenamefont {Weniger}(1989)}]{Weniger:89}%
  \BibitemOpen
  \bibfield  {author} {\bibinfo {author} {\bibfnamefont {E.~J.}\ \bibnamefont
  {Weniger}},\ }\href@noop {} {\bibfield  {journal} {\bibinfo  {journal}
  {Comput. Phys. Rep.}\ }\textbf {\bibinfo {volume} {10}},\ \bibinfo {pages}
  {189} (\bibinfo {year} {1989})}\BibitemShut {NoStop}%
\bibitem [{\citenamefont {\'Cwiok}\ \emph {et~al.}(1994)\citenamefont
  {\'Cwiok}, \citenamefont {Jeziorski}, \citenamefont {Ko\l{}os}, \citenamefont
  {Moszy\'nski},\ and\ \citenamefont {Szalewicz}}]{Cwiok1994symmetry}%
  \BibitemOpen
  \bibfield  {author} {\bibinfo {author} {\bibfnamefont {T.}~\bibnamefont
  {\'Cwiok}}, \bibinfo {author} {\bibfnamefont {B.}~\bibnamefont {Jeziorski}},
  \bibinfo {author} {\bibfnamefont {W.}~\bibnamefont {Ko\l{}os}}, \bibinfo
  {author} {\bibfnamefont {R.}~\bibnamefont {Moszy\'nski}}, \ and\ \bibinfo
  {author} {\bibfnamefont {K.}~\bibnamefont {Szalewicz}},\ }\href {\doibase
  10.1016/0166-1280(94)80124-X} {\bibfield  {journal} {\bibinfo  {journal} {J.
  Mol. Struct. THEOCHEM}\ }\textbf {\bibinfo {volume} {307}},\ \bibinfo {pages}
  {135} (\bibinfo {year} {1994})}\BibitemShut {NoStop}%
\bibitem [{\citenamefont {\'Cwiok}\ \emph {et~al.}(1992)\citenamefont
  {\'Cwiok}, \citenamefont {Jeziorski}, \citenamefont {Ko\l{}os}, \citenamefont
  {Moszy\'nski}, \citenamefont {Rychlewski},\ and\ \citenamefont
  {Szalewicz}}]{Cwiok:92:Pol}%
  \BibitemOpen
  \bibfield  {author} {\bibinfo {author} {\bibfnamefont {T.}~\bibnamefont
  {\'Cwiok}}, \bibinfo {author} {\bibfnamefont {B.}~\bibnamefont {Jeziorski}},
  \bibinfo {author} {\bibfnamefont {W.}~\bibnamefont {Ko\l{}os}}, \bibinfo
  {author} {\bibfnamefont {R.}~\bibnamefont {Moszy\'nski}}, \bibinfo {author}
  {\bibfnamefont {J.}~\bibnamefont {Rychlewski}}, \ and\ \bibinfo {author}
  {\bibfnamefont {K.}~\bibnamefont {Szalewicz}},\ }\href {\doibase
  10.1016/0009-2614(92)85912-T} {\bibfield  {journal} {\bibinfo  {journal}
  {Chem. Phys. Lett.}\ }\textbf {\bibinfo {volume} {195}},\ \bibinfo {pages}
  {67} (\bibinfo {year} {1992})}\BibitemShut {NoStop}%
\bibitem [{\citenamefont {Jeziorski}\ \emph {et~al.}(1978)\citenamefont
  {Jeziorski}, \citenamefont {Szalewicz},\ and\ \citenamefont
  {Cha\l{}asi\'nski}}]{Jeziorski:78}%
  \BibitemOpen
  \bibfield  {author} {\bibinfo {author} {\bibfnamefont {B.}~\bibnamefont
  {Jeziorski}}, \bibinfo {author} {\bibfnamefont {K.}~\bibnamefont
  {Szalewicz}}, \ and\ \bibinfo {author} {\bibfnamefont {G.}~\bibnamefont
  {Cha\l{}asi\'nski}},\ }\href {\doibase 10.1002/qua.560140306} {\bibfield
  {journal} {\bibinfo  {journal} {Int. J. Quant. Chem.}\ }\textbf {\bibinfo
  {volume} {14}},\ \bibinfo {pages} {271} (\bibinfo {year} {1978})}\BibitemShut
  {NoStop}%
\bibitem [{\citenamefont {{\'C}wiok}\ \emph {et~al.}(1992)\citenamefont
  {{\'C}wiok}, \citenamefont {Jeziorski}, \citenamefont {Ko\l{}os},
  \citenamefont {Moszy\'nski},\ and\ \citenamefont {Szalewicz}}]{Cwiok:92:SRS}%
  \BibitemOpen
  \bibfield  {author} {\bibinfo {author} {\bibfnamefont {T.}~\bibnamefont
  {{\'C}wiok}}, \bibinfo {author} {\bibfnamefont {B.}~\bibnamefont
  {Jeziorski}}, \bibinfo {author} {\bibfnamefont {W.}~\bibnamefont {Ko\l{}os}},
  \bibinfo {author} {\bibfnamefont {R.}~\bibnamefont {Moszy\'nski}}, \ and\
  \bibinfo {author} {\bibfnamefont {K.}~\bibnamefont {Szalewicz}},\ }\href
  {\doibase 10.1063/1.463475} {\bibfield  {journal} {\bibinfo  {journal} {J.
  Chem. Phys.}\ }\textbf {\bibinfo {volume} {97}},\ \bibinfo {pages} {7555}
  (\bibinfo {year} {1992})}\BibitemShut {NoStop}%
\bibitem [{\citenamefont {{T. C. Scott, J. F. Babb, A. Dalgarno, and J. D.
  Morgan III}}(1993)}]{Scott:93}%
  \BibitemOpen
  \bibfield  {author} {\bibinfo {author} {\bibnamefont {{T. C. Scott, J. F.
  Babb, A. Dalgarno, and J. D. Morgan III}}},\ }\href {\doibase
  10.1063/1.465193} {\bibfield  {journal} {\bibinfo  {journal} {J. Chem.
  Phys.}\ }\textbf {\bibinfo {volume} {99}},\ \bibinfo {pages} {2841} (\bibinfo
  {year} {1993})}\BibitemShut {NoStop}%
\bibitem [{\citenamefont {Bartlett}(1955)}]{Bartlett:55}%
  \BibitemOpen
  \bibfield  {author} {\bibinfo {author} {\bibfnamefont {J.~H.}\ \bibnamefont
  {Bartlett}},\ }\href {\doibase 10.1103/PhysRev.98.1067} {\bibfield  {journal}
  {\bibinfo  {journal} {Phys. Rev.}\ }\textbf {\bibinfo {volume} {98}},\
  \bibinfo {pages} {1067} (\bibinfo {year} {1955})}\BibitemShut {NoStop}%
\end{thebibliography}%

\end{document}